\documentclass[conference]{IEEEtran}

\usepackage{amsmath}
\usepackage{xfrac}
\usepackage{amsmath,amssymb,amsfonts}
\usepackage{graphicx}
\usepackage{array}
\usepackage{makecell}

\usepackage{tikz}
\usepackage{pgfplots}
\usepackage{subfigure}	
\usepackage{xcolor}
\usetikzlibrary{arrows,shadows,petri}

\usepackage{enumerate}

\usepackage{booktabs}
\usepackage{floatflt}
\usepackage{enumerate}
\usepackage{psfrag}	
\usepackage{array}
\usepackage{multirow,hhline}
\usepackage{exscale}
\usepackage{color}
\usepackage{colortbl}
\usepackage{epsfig,subfigure}
\usepackage{cite}
\usepackage{amsthm}
\usepackage[font=small]{caption}
\usepackage[latin1]{inputenc} 
\usepackage[T1]{fontenc} 
\usepackage{enumerate}
\usepackage{balance}

\usetikzlibrary{arrows}
\usepackage{marvosym}


\newtheorem{lemma}{Lemma}

\usetikzlibrary{shapes,snakes}
\usetikzlibrary{calc}
\usetikzlibrary{patterns}
\usetikzlibrary{decorations.pathmorphing}

\newcommand{\antennaTx}[3]{
\coordinate (a) at (#1,#2);
\draw[scale=(#3)] ($(a)$) -- ($(a)+(0.1,0)$) -- ($(a)+(0.1,0.2)$) -- ($(a)+(0.07,0.23)$) -- ($(a)+(0.13,0.23)$) -- ($(a)+(0.1,0.2)$);
}

\begin{document}

\IEEEoverridecommandlockouts
\title{Optimal Resource Allocation for Joint Sensing and Communication: Multiple Targets and Clutters}
\author{
\IEEEauthorblockN{Ali Kariminezhad, \textit{Student Member, IEEE}, Soheil Gherekhloo, \textit{Member, IEEE},\\ and Aydin Sezgin, \textit{Senior Member, IEEE}}\\
\thanks{
A. Kariminezhad and A. Sezgin are with the Institute of Digital Communication Systems, Ruhr-Universit\"at Bochum (RUB), Germany (emails: \{ali.kariminezhad, soheyl.gherekhloo, aydin.sezgin\}@rub.de). S. Gherekhloo is with the Chassis Systems Control, Robert Bosch GmbH, Heilbronn, Germany (email: Soheil.Gherekhloo@de.bosch.com) 
}}
\maketitle

\begin{abstract}
We study a contactless target probing based on stimulation by a radio frequency (RF) signal. The transmit signal is dispatched from a transmitter equipped with a two-dimensional antenna array. Then, the reflected signal from the targets are received at multiple distributed sensors. The observation at the sensors are amplified and forwarded to the fusion center. Afterwards, the fusion center performs space-time post processing to extract the maximum common information between the received signal and the targets impulse responses. Optimal power allocation at the transmitter and amplification at the sensors is investigated. The sum-power minimization problem turns out to be a non-convex problem. We propose an efficient algorithm to solve this problem iteratively. By exploiting maximum-ratio transmission (MRT), maximum-ratio combining (MRC) of space-time received signal vector is the optimal receiver at sufficiently low signal-to-interference-plus-noise-ratio (SINR). However, zero-forcing (ZF) at the fusion center outperforms MRC at higher SINR demands. 
\end{abstract}

\section{Introduction}
Environment sensing for the aim of targets' material characterization requires efficient system design and parameter estimation. Targets to be identified can be classified into two categories, namely, active signal sources and passive objects. Assuming active signal sources, the signals are detected by multiple sensors, which is then forwarded to a fusion center for joint processing. The authors in~\cite{Alirezaei2014,Wang2011} provide analytical solution for the power allocation at the sensors under sensor sum-power and individual power constraint. Furthermore, a similar system is investigated in~\cite{Godrich2011,Shen2012,Guo2017,Zhang2016} from different perspectives including localization, scheduling and energy harvesting. In contrary to active signal sources, passive targets are silent, hence they require stimulation by external signal sources to be activated. The systems that deal with passive targets are known as radar systems. These systems can be designed to be collocated, i.e., the transmitter and received are embedded in a single device. The authors in~\cite{Yang2007} study the optimal waveform design in a collocated multi-input multi-output (MIMO) system with a single extended target. In that work, the mutual information between the transmit and received signals is maximized. Moreover, they study the waveform design for minimizing the mean-squared of the channel estimation error. Moreover, the authors in~\cite{Leshem2007} study similar problem assuming multiple extended targets. For distributed transceivers, the authors in~\cite{Jeong2016} study the optimal waveform design that maximizes the so-called Bhattacharyya distance. That work mainly focuses on single-target detection in a single-clutter environment.

\begin{figure}
\centering
\tikzset{every picture/.style={scale=1}, every node/.style={scale=0.7}}
\begin{tikzpicture}[scale=0.8,
    knoten/.style={
      circle,
      inner sep=.35cm,
      draw},
    ]
\draw[fill=yellow!20] (-0.7,-2) rectangle (0.7,2);
\antennaTx{0.7}{1.6}{2};
\antennaTx{0.7}{0.7}{2};
\node at (0.85,0.2){.};
\node at (0.85,-0.3){.};
\node at (0.85,-0.8){.};
\antennaTx{0.7}{-1.7}{2};
\draw[decorate,decoration=expanding waves,segment length=1mm,segment angle=10,scale=1] (1.2,0) -- (2.4,0.8);
\draw[decorate,decoration=expanding waves,segment length=1mm,segment angle=10,scale=1] (1.2,0) -- (3.3,-0.2);
\node[scale=0.4, shading=ball, ball color=yellow!80!black] at (3,1.2) (knoten1) [knoten] {1};

\draw[decorate,decoration=expanding waves,segment length=1mm,segment angle=360,scale=1] (3,1.2) -- (3.7,1.2);

\node[scale=0.4, shading=ball, ball color=red!80!black] at (4,-0.2) (knoten2) [knoten] {2};

\draw[decorate,decoration=expanding waves,segment length=1mm,segment angle=360,scale=1] (4,-0.2) -- (4.7,-0.2);

\node[scale=0.4, shading=ball, ball color=cyan!80!black] at (3.5,-1.5) (knoten3) [knoten] {3};

\draw[decorate,decoration=expanding waves,segment length=1mm,segment angle=360,scale=1] (3.5,-1.5) -- (3.9,-1.5);

\draw[fill=black] (5.5,1.8) circle (0.08cm);
\antennaTx{5.9}{1.8}{1};
\draw[fill=black] (5.5,0.9) circle (0.08cm);
\antennaTx{5.9}{0.9}{1};
\draw[fill=black] (5.5,-1.8) circle (0.08cm);
\antennaTx{5.9}{-1.8}{1};
\node at (5.7,0.2){.};
\node at (5.7,-0.3){.};
\node at (5.7,-0.8){.};
\draw[fill=yellow!20] (5.5,-2) rectangle (5.9,-1.6);
\draw[fill=yellow!20] (5.5,0.7) rectangle (5.9,1.1);
\draw[fill=yellow!20] (5.5,1.6) rectangle (5.9,2);

\node [fill=yellow!20,cloud, draw,cloud puffs=15,cloud puff arc=90, aspect=2, inner ysep=1.5em,rotate=90] at (8,0){Fusion Center};
\end{tikzpicture}
\caption{RF-based stimulation of the objects in the sensing environment with amplify-and-forward sensors. The two top objects are of targets and the bottom one is clutter.}
\label{fig:sensorNet}
\end{figure}
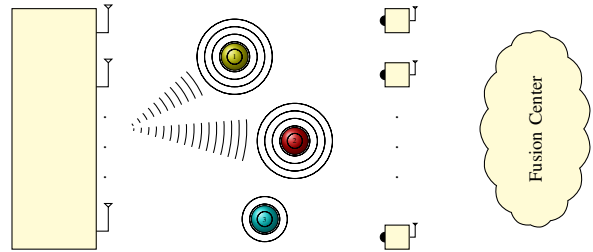
Having multiple targets and clutters in the sensing environment, in this paper, a radar system is exploited for material characterization purposes. This characterization can be fulfilled by estimating the second-order moment of the materials. We address this radar system system by utilizing a two-dimensional multi-antenna transmitter which pre-processes the signal given the position of the objects in the sensing environment. This design allows three-dimensional beamforming which has been shown to be beneficial~\cite{Razavizadeh2014,Nam2013,Koppenborg2012,Karaman2009,Dhanantwari2004}. Here, we assume that the position of the objects are known. Having this information by, for instance, the methods in~\cite{Gezici2005} and the references therein for localization problems, the dispatched signal power at the objects surface is maximized by transmitting in the direction of their steering vectors. The incident signal in then reflected by the objects in the environment. Sensing the response of the objects in a particular spectrum can help classifying the material of the objects. For instance, in photo-acoustic imaging, the target is stimulated at a very high frequency spectrum, however the response is captured at the ultrasonic frequency range~\cite{Francis2014,Qu2010,Naam2015}. However, the reflection by the targets over the same transmit signal spectrum can also be helpful for identification purposes. Therefore at the same frequency spectrum, the reflected signal from the objects are sensed, amplified at multiple sensors, and then forwarded to the central processing unit which is referred to as fusion center. The fusion center is equipped with multiple single-antenna baseband units with high capacity links. The receive antennas at the fusion center observe the noisy version of the forwarded signals from the sensors,~Fig.~\ref{fig:sensorNet}. Then, the fusion center performs post-processing for target response detection. For this purpose, we minimize the transmit power and sensor amplification under received signal-to-interference-plus-noise ratio (SINR) constraint. This problem turns out to be a signomial program, which is essentially a non-convex problem. We exploit an efficient algorithm to obtain a good sub-optimal solution iteratively.

\section{System Model}
We consider a sensing environment with $N$ targets of interest and $N^{'}$ clutters, i.e., $N+N^{'}$ objects in total. The RF signal from a single multi-antenna transmitter with a $M\times M^{'}$ planner antenna array stimulates the objects in the sensing environment, where there is a line-of-sight (LoS) between the transmitter and objects. We assume that, the transmit antennas are equi-distantly (uniform linear array) positioned in two-dimensional Cartesian basis dimensions. Here we consider the antenna at the center of coordinates as the reference antenna. By assuming half-wavelength distance between horizontal and vertical antenna elements, we obtain the steering vector corresponding to the object $i$ as
\begin{align}
\mathbf{a}_i=\big[1,&\cdots,e^{j\pi( m\sin\theta\sin\phi+m^{'}\cos\phi)},\cdots\nonumber\\
&,e^{j\pi( (M-1)\sin\theta\sin\phi+(M^{'}-1)\cos\phi)}\big]^{T}\in\mathcal{C}^{MM^{'}},
\end{align}
where $m\in\{0,\cdots,M-1\}$ and $m^{'}\in\{0,\cdots,M^{'}-1\}$.
Notice that the azimuth and elevation angles of the object $i$ are represented by $\theta_i$ and $\phi_i$, respectively. Now, having the steering vectors of the targets, the transmit signal is formed as
$
\mathbf{s}=\sum_{j=1}^{N}\mathbf{u}_jd_j,
$
where $\mathbf{u}_j\in\mathcal{C}^{MM^{'}}$ specifies the beam direction towards $j$th target and $d_j$ is the transmit symbol to the $j$ target. The corresponding allocated power to the $j$th target is represented by $p_j=\mathbb{E}\{|d_j|^2\}$. Now, the reflected  signals from the objects are given by
\begin{align}
x_i=\mathbf{a}^{H}_i\mathbf{s}l_i,\ \forall i\in\{\mathcal{N}\cup\mathcal{N}^{c}\},
\end{align} 
where $\mathcal{N}$ and its complement $\mathcal{N}^{c}$ are the sets of targets and clutters, respectively, i.e.,  $\mathcal{N}=\{1,...,N\}$ and $\mathcal{N}^{c}=\{N+1,...,N+N^{'}\}$. Moreover, the response of object $i$ for the incident RF signal is represented by $l_i$. Here, we assume that this response is a random variable with Gaussian distribution. Therefore, estimating the second-order moment $\mathbb{E}\{|l_i|^{2}\},\ \forall i$ helps classifying the objects.
Now the signal $x_i,\ \forall i$ is sensed by $K$ distributed sensors, see Fig.~\ref{fig:sensorNet}. Then, the received signal at the $k$th sensor is given by
\begin{align}
y_k=\sum_{i=1}^{N+N^{'}}g_{ik}x_i+n_k,\ \forall k\in\mathcal{K}=\{1,...,K\},
\end{align}
where $g_{ik}\in\mathbb{C}$ is the channel from object $i$ towards sensor $k$ and $n_k\in\mathbb{C}$ is the additive receiver noise at the sensor $k$. Here, we assume that the noise at all sensors follow identical and independent  zero-mean Gaussian distribution. Moreover, the noise variance at the sensor $k$ is given by $\sigma^{2}_k$. The signal is amplified at the sensors and forwarded to the fusion center in different time slots. Then, the post-processed received signal at the fusion center over $R$ antennas and $K$ time instants is written as
\begin{align}
\mathbf{z}=\mathbf{V}^{H}\left(\sum_{i=1}^{N+N^{'}} \sqrt{\delta_i}l_i\mathbf{w}_i+\mathbf{n}^{'} \right),\label{fusionCenter}
\end{align}\\
where $\mathbf{V}\in\mathbb{C}^{KR\times N}$ is the post processing matrix. Notice that $\mathbf{V}=[\mathbf{v}_1,\ ...,\mathbf{v}_N]$, where $\mathbf{v}_j,\ \forall j\in\mathcal{N}$, is the $j$th target post processing vector. Moreover, the power of the signal received at target $i$ is represented by $\delta_i=\mathbb{E}\{|\mathbf{a}^{H}_i \mathbf{s}|^2\}$. The vectors $\mathbf{w}_i$ and $\mathbf{n}^{'}$ are the equivalent channel and noise vectors which are defined as
\begin{align}
\mathbf{w}_i&=\begin{bmatrix}
\sqrt{\alpha_1}g_{i1}\mathbf{f}^{T}_1 &...& \sqrt{\alpha_K}g_{iK}\mathbf{f}^{T}_K
\end{bmatrix}^{T},\label{wi}\\
\mathbf{n}^{'}&=\underbrace{\begin{bmatrix}
\sqrt{\alpha_1}n_1\mathbf{f}^{T}_1 &...& \sqrt{\alpha_K}n_K\mathbf{f}^{T}_K
\end{bmatrix}^{T}}_{\mathbf{n}_{\text{s}}}+\mathbf{n}_{\text{fc}},\label{nPrime}
\end{align}
where $\alpha_k$ is the amplification factor at the $k$th sensor and $\mathbf{f}_k\in\mathbb{C}^{R},\forall k\in\mathcal{K}$ is the communication channel from $k$th sensor to the fusion center. Note that, $\mathbf{n}_{\text{fc}}\in\mathbb{C}^{KR}$ is the zero-mean Gaussian noise at the fusion center with $R$ antennas over $K$ time slots with the covariance matrix $\sigma^{2}_{\text{fc}}\mathbf{I}$.

In what follows, we assume that the steering vectors corresponding with the targets are known at the transmitter, i.e., $a_i,\ \forall i\in\mathcal{N}$. Whereas, the following knowledge is known at the fusion center, 1) $a_i\ \forall i\in\mathcal{N}\cup\mathcal{N}^{c}$, 2) $g_{ik},\ \forall i\in\mathcal{N}\cup\mathcal{N}^{c}, k\in\mathcal{K}$, 3) $\mathbf{f}_k,\ \forall k\in\mathcal{K}$. In the next sections we discuss the pre- and post-processing schemes exploited at the transmitter and the fusion center, respectively.

\subsection{Pre-processing}\label{sec:PreProc}
Given the steering vectors of the targets at the transmitter, maximum ratio transmission (MRT) is the optimal scheme. In MRT, the transmit directions toward the $j$th target ($j\in\mathcal{N}$) is adjusted to the corresponding steering vectors, i.e., $\mathbf{a}_j$. Hence,
\begin{align}
\mathbf{u}_j=\frac{\mathbf{a}_j}{\|\mathbf{a}_j\|}=\frac{1}{\sqrt{MM^{'}}}\mathbf{a}_j.\label{Mrt1}
\end{align}
Utilizing this filter at the transmitter the received signal power at $i$th object (either a target or a clutter) is written as
\begin{align}
\delta_j= & MM^{'}p_j+\sum_{\substack{i=1\\i\neq j}}^{N}p_i\mathbf{a}^{H}_j\mathbf{u}_i
\mathbf{u}^{H}_i\mathbf{a}_j\quad j\in\mathcal{N},\label{pPrime1}\\
\delta_j= &\sum_{i=1}^{N} p_i\mathbf{a}^{H}_j\mathbf{u}_i
\mathbf{u}^{H}_i\mathbf{a}_j,\quad\forall j\in\mathcal{N}^{c},\label{pPrime2}
\end{align}
where $MM^{'}$ is the antenna gain for the $j$th target.

\subsection{Post-processing}
As can be noticed from~\eqref{fusionCenter}, the post-processed signal includes both desired and interference components. This can be seen by
\begin{align}
z_j=\mathbf{v}_j^{H}(\underbrace{\sqrt{\delta_j}l_j\mathbf{w}_j}_{\text{desired}}+ \underbrace{\sum_{\substack{i=1\\i\neq j}}^{N+N^{'}}\sqrt{\delta_i}l_i\mathbf{w}_i}_{\text{interference}} +\mathbf{n}^{'} ),\ \forall j\in\mathcal{N},\label{fusionCenter2}
\end{align}
where the $j$th column of the post-processing matrix $\mathbf{V}$ is denoted by $\mathbf{v}_j\in\mathbb{C}^{KR}$. Now, the post-processing filters, power allocation per target and signal amplification at the sensors need to be designed to guarantee a certain threshold is differentiating the targets $l_j,\forall j\in\mathcal{N}$. Here, mutual information is exploited as the information measure as
$
I(z_j;l_j)=\log_2(1+\rho_j),\quad\forall j\in\mathcal{N},
$
where $\rho_j$ is the SINR corresponding with $j$th target. Here, mutual information is a monotonically increasing function in $\rho_j$.
We formulate the SINR for target $j$ as
$
\rho_j=\frac{\delta_jQ_j\mathbf{v}^{H}_j\mathbf{w}_j\mathbf{w}^{H}_j\mathbf{v}_j}
{\Sigma_{j_\text{int}}+\Sigma_{j_{n_{\text{s}}}}+\Sigma_{j_{n_{\text{fc}}}}},\quad\forall j\in\mathcal{N},
$
where the interference and noise variances are
\begin{align}
\Sigma_{j_\text{int}}&=\mathbf{v}^{H}_j\sum_{\substack{i=1\\ i\neq j}}^{N+N^{c}}\delta_iQ_i
\mathbf{w}_i\mathbf{w}^{H}_i\mathbf{v}_j\\
\Sigma_{j_{n_{\text{s}}}}&=\mathbf{v}^{H}_j\mathbf{A}_n\mathbf{v}_j,\quad
\Sigma_{j_{n_{\text{fc}}}}=\sigma^{2}_{\text{fc}}||\mathbf{v}_j||^{2},
\end{align}
respectively. The equivalent sensor noise covariance matrix observed in decoding the information of the $j$th target is denoted by $\mathbf{A}_n\in\mathbb{C}^{KR\times KR}$ which is a block diagonal matrix with the $k$th block represented by $\alpha_k\sigma^{2}_n\mathbf{f}_k\mathbf{f}^{H}_k$.\\
Now, given MRT at the transmitter, we consider following signal combining strategies at the fusion center,
A) maximum-ratio combining (MRC): maximizes signal-to-noise ratio (SNR).
B) zero-forcing (ZF): maximizes signal-to-interference ratio (SIR).
In what follows we discuss these schemes.
\subsubsection{Maximum-ratio combining}
Assuming MRC at the fusion center, the following signal combining vector maximizes SNR,
$
\mathbf{v}^{(\text{MR})}_{j}=\mathbf{w}_j\quad\forall j\in\mathcal{N},
$
which is less complex for practical implementations, however does not consider the destructive effect of interference in the signal combining phase. Utilizing MRC, we will minimize the sum transmit power plus sum power amplification at the sensors jointly.

\subsubsection{Zero-forcing}
Here, we enforce the interference to zero while decoding the signal of $j$th target. This can be done in space-time by
$
\mathbf{v}^{(\text{ZF})}_{j}=\text{null}\{\mathbf{w}_1,...,\mathbf{w}_{j-1},\mathbf{w}_{j+1},...,\mathbf{w}_{N+N^{c}}\}\label{Zf1}
$, where ZF combining vector spans the null-space of the interference dimensions. However, the optimal combining vector in this null-space for the $j$th target is the $j$th column of
$
\mathbf{V}^{(\text{ZF})^{\star}}=\mathbf{W}
\left(\mathbf{W}^{H}\mathbf{W}\right)^{-1},
$
where $\mathbf{W}=\begin{bmatrix} \mathbf{w}_1, ... ,  \mathbf{w}_{N+N^{c}} \end{bmatrix}$.

\section{Optimization Problem}\label{sec:optimizationProblem}
In this section, we formulate sum-power minimization problems under target SINR constraints. Exploiting maximum-ratio transmission (MRT) at the transmitter and maximum-ratio combining (MRC) at the fusion center, the sum transmit power plus sum power amplification minimization problem is formulated as
\begin{subequations}\label{A1}
\begin{align}
\hspace*{-2cm}\min_{p_j,\alpha_k,\ \forall j,k}\quad & \sum_{j=1}^{N}p_j+\sum_{k=1}^{K}\alpha_k \tag{\ref{A1}}\\
\text{subject to}\quad & \rho_j\geq \psi_j,\label{A11}\\
& \sum_{j=1}^{N}p_j\leq P_{\text{max}},\quad
 \alpha_k\leq \alpha_{\text{max}},
\end{align}
\end{subequations}
where the $j$th target SINR demand is defined by $\psi_j$. Furthermore, the sum transmit power is restricted by $P_{\text{max}}$ and the maximum amplification power of each sensor is limited by $\alpha_{\text{max}}$, respectively. Evidently, the objective function is affine, however SINR constraints in~\eqref{A11} produce a non-convex set. Utilizing MRT at the transmitter, the SINR expression for the $j$th target is written as
\begin{align}
\rho_j=\frac{ \Sigma_{j_\text{des}} }{ \Sigma_{j_\text{int}} + \Sigma_{j_{n_{\text{s}}}} + \Sigma_{j_{n_{\text{fc}}}} },
\end{align}
where
\begin{align}
\Sigma_{j_\text{des}}&=\delta_j Q_j  \left(\sum_{k=1}^{K} \alpha_k|g_{jk}|^{2}\| \mathbf{f}_k \|^{2}\right)^{2},\label{des}\\
\Sigma_{j_\text{int}}&=\sum_{i\neq j}\delta_i Q_i \left| \sum_{k=1}^{K} \alpha_kg_{jk}g^{*}_{ik}\|\mathbf{f}_k\|^{2}\right|^{2},\label{int}\\
\Sigma_{j_{n_{\text{s}}}}&=\sum_{k=1}^{K}\sigma^{2}_{n_k}\alpha^{2}_k|g_{jk}|^{2}\|\mathbf{f}_k\|^{4},\label{ns}\\
\Sigma_{j_{n_{\text{fc}}}}&=\sigma^{2}_{\text{fc}} \sum_{k=1}^{K} \alpha_k|g_{jk}|^{2}\| \mathbf{f}_k \|^{2}.\label{nfc}
\end{align}
Here, we assume the thermal noise variance at the fusion center and sensors are equal, i.e., $\sigma^2=\sigma_\text{fc}=\sigma_{n_k},\ \forall k$. Having the SINR for the $j$th target, the constraint~\eqref{A11} can be reformulated as
\begin{align}
\frac{\psi_j\left( \Sigma_{j_\text{int}} + \Sigma_{j_{n_{\text{s}}}} + \Sigma_{j_{n_{\text{fc}}}}\right)}{\Sigma_{j_\text{des}}}\leq 1,\ \forall j\in\mathcal{N}\label{INSR}
\end{align}
where $\Sigma_{j_\text{des}}$, $\Sigma_{j_{n_{\text{s}}}}$ and  $\Sigma_{j_{n_{\text{fc}}}}$ are posynomials and $\Sigma_{j_\text{int}}$ is a signomial, in general.
\begin{lemma}
$\Sigma_{j_\text{int}}$ is a posynomial if the following constraint holds, ($\forall i\neq j\ \text{and}\ \forall k\neq l,\forall\psi\in\mathbb{Z}$)
\begin{align}
&2\psi\pi-\frac{\pi}{2}\leq\cos\left(\measuredangle g_{jk}g^{*}_{jl}g^{*}_{ik}g_{il}\right)\leq 2\psi\pi+\frac{\pi}{2},
\end{align}
\begin{proof}
The expression in~\eqref{int} is rewritten as
\begin{align}
\sum_{i\neq j}\delta_i Q_i \left( \underbrace{\sum_{k=1}^{K} \alpha_k\|\mathbf{f}_k\|^{2}g_{jk}g^{*}_{ik}\sum_{l=1}^{K} \alpha_l\|\mathbf{f}_l\|^{2}g^{*}_{jl}g_{il}}_{\Gamma(\boldsymbol{\alpha)}} \right),
\end{align}
where we define the expression in the braces as $\Gamma(\boldsymbol{\alpha})$, with $\boldsymbol{\alpha}=[\alpha_1,\cdots,\alpha_K]$. Notice that, $\Gamma(\boldsymbol{\alpha})$ is the summation of $K^2$ monomial functions. The monomials corresponding with $k=l$ have real positive values, since $g_{jk}g^{*}_{ik}g_{jl}^{*}g_{il}=|g_{jk}|^2|g_{ik}|^2\geq 0$. The monomials corresponding with $k\neq l$ are
\begin{align}
&\alpha_k\alpha_l\|\mathbf{f}_k\|^{2}\|\mathbf{f}_l\|^{2}
\left(g_{jk}g^{*}_{ik}g_{jl}^{*}g_{il}+g^{*}_{jk}g_{ik}g_{jl}g_{il}^{*}\right),\nonumber\\
&=2\alpha_k\alpha_l\|\mathbf{f}_k\|^{2}\|\mathbf{f}_l\|^{2}\Re\{g_{jk}g^{*}_{ik}g_{jl}^{*}g_{il}\},
\label{Lemma1A}
\end{align} 
which do not necessarily yield a positive value. Hence, $\Sigma_{j_\text{int}}$ is a signomial function in $p_j\ \forall j\in\mathcal{N}^{t}$ and $\alpha_k,\ k\in\mathcal{K}$. The expression~\eqref{Lemma1A} is positive if
\begin{align}
&2\psi\pi-\frac{\pi}{2}\leq\cos\left(\measuredangle g_{jk}g^{*}_{jl}g^{*}_{ik}g_{il}\right)\leq 2\psi\pi+\frac{\pi}{2},\ \ \forall\psi\in\mathbb{Z}
\end{align}
\end{proof}
\end{lemma}
\begin{figure*}[ht]%
\centering
\subfigure[]{
\tikzset{every picture/.style={scale=1}, every node/.style={scale=.9}}%
\begin{tikzpicture}
\begin{axis}[%
xmin=1,
xmax=12,
xtick={1,3,5,7,9,11},
xlabel={$q$ (iteration index)},
xmajorgrids,
ymin=11.6,
ymax=13,
ytick={11.6,11.8,12,12.2,12.4,12.6,12.8,13},
ylabel={$\min\ \sum_{j=1}^{N}p_j+\sum_{k=1}^{K}\alpha_k$ [dB]},
ymajorgrids,
legend style={at={(axis cs: 0,0)},anchor=north east,draw=black,fill=white,legend cell align=left}
]
\addplot [color=blue,solid,mark=asterisk, mark options={solid}]
  table[row sep=crcr]{1	13\\
    2	12.4409812802194\\
    3	12.1158639285995\\
    4	11.9355328970433\\
    5	11.8367005567424\\
    6	11.7826050972283\\
    7	11.7527980141098\\
    8	11.7361515414647\\
    9	11.7266657564504\\
    10	11.7211124508479\\
    11	11.7177503908635\\
    12	11.7156345177447\\
    };
\end{axis}
\end{tikzpicture}%
\label{Res:Converg1}
}\quad
\subfigure[]{
\tikzset{every picture/.style={scale=1}, every node/.style={scale=.9}}%
\begin{tikzpicture}
\begin{axis}[%
xmin=0.01,
xmax=1.6,
xtick={0,0.2,0.4,0.6,0.8,1,1.2,1.4,1.6},
xlabel={$\psi$ (SINR demands)},
xmajorgrids,
ymin=-8,
ymax=22,
ytick={-8,-4,0,4,8,12,16,20},
ylabel={$\min\ \sum_{j=1}^{N}p_j+\sum_{k=1}^{K}\alpha_k$ [dB]},
ymajorgrids,
legend style={at={(axis cs: 1.6,-8)},anchor=south east,draw=black,fill=white,legend cell align=left}
]
\addplot [color=red,solid,mark=o]
  table[row sep=crcr]{0.001	7.78288253596396\\
  0.101	7.93091874560761\\
  0.201	8.10468796184524\\
  0.301	8.31172392130254\\
  0.401	8.56291302170422\\
  0.501	8.87466266446007\\
  0.601	9.27307593292113\\
  0.701	9.80280708517881\\
  0.801	10.5485697228352\\
  0.901	11.6993467951679\\
  1.001	13.8256681863065\\
  1.101	21.6216005675223\\
  };
\addlegendentry{\small opt. Tx-power, max. amp., MRC};
\addplot [color=blue,solid]
  table[row sep=crcr]{0.001	7.79462450122965\\
  0.101	8.8259386928851\\
  0.201	9.65967305981543\\
  0.301	10.3588393398409\\
  0.401	10.960900076714\\
  0.501	11.4895685829242\\
  0.601	11.9608095567636\\
  0.701	12.3858837757046\\
  0.801	12.7730386683028\\
  0.901	13.1284875481136\\
  1.001	13.4570327280914\\
  1.101	13.7624618502251\\
  1.201	14.0478150240583\\
  1.301	14.3155694543259\\
  1.401	14.5677703685763\\
  1.501	14.8061259615981\\
  1.601	15.0320776080934\\
  };
\addlegendentry{\small opt. Tx-power, max. amp., ZF};

\addplot [color=green!50!black,solid,mark=*]
  table[row sep=crcr]{0.01	-7.27384572351192\\
    0.06	-2.69862701130313\\
    0.11	-0.95147815995086\\
    0.16	0.218617659449915\\
    0.21	1.14444635005603\\
    0.26	1.92865020042462\\
    0.31	2.62483160032072\\
    0.36	3.26336255053232\\
    0.41	3.863619423597\\
    0.46	4.43920488428394\\
    0.51	5.00039103572207\\
    0.56	5.55502635300853\\
    0.61	6.11191953250629\\
    0.66	6.67872529767108\\
    0.71	7.26200817589618\\
    0.76	7.87512648997187\\
    0.81	8.52412884941269\\
    0.86	9.23048582412233\\
    0.91	10.0095303076987\\
    0.96	10.8966939595631\\
    1.01	11.9418696887074\\
    1.06	13.2374269788138\\
    1.11	14.9621431401173\\
    1.11	21.6216005675223\\
    };
\addlegendentry{\small opt. Tx-power, opt. amp. (Joint opt.), MRC};

\end{axis}
\end{tikzpicture}%
\label{Res:MRC}
}
\caption{(a) Convergence of the minimum sum transmit power plus sum sensor amplification factor. The SINR demands for the targets is set to $1$ and assumed to be equal. (b) Minimum required power to fulfill the SINR constraints of the targets. the SINR constraint are assumed to be equal.}
\end{figure*}
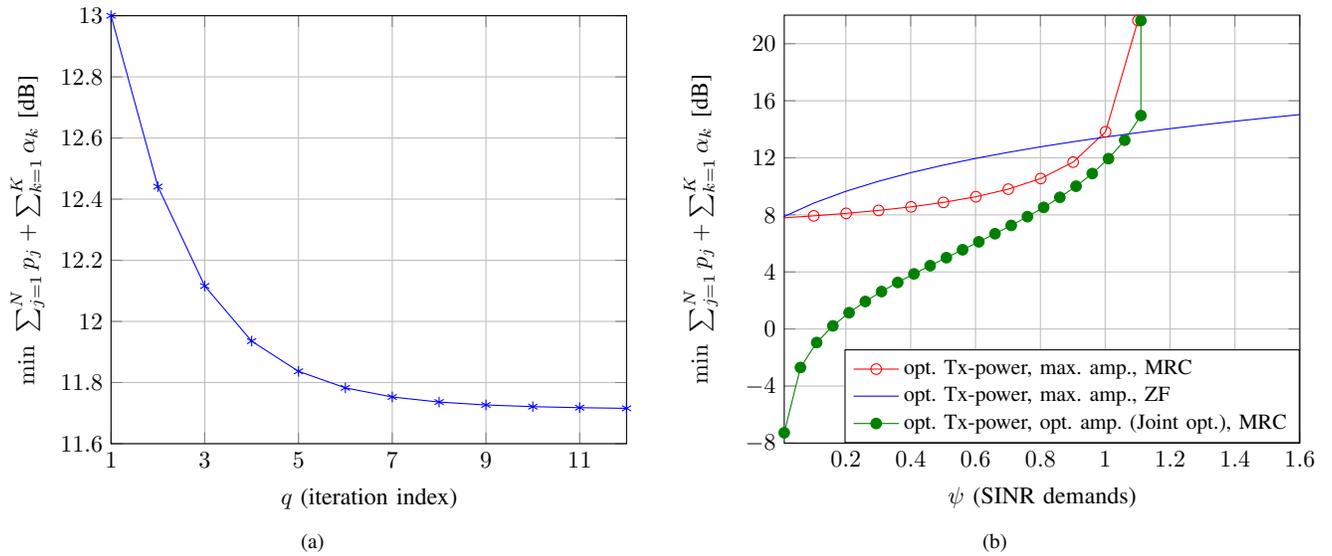 
Here, we consider the general case where $\Sigma_{j_\text{int}}$ is a signomial. In general, the expression in~\eqref{int} can be reformulated as the difference of two posynomials as
$
\Sigma_{j_\text{int}}=\Sigma^{(1)}_{\text{int}}-\Sigma^{(2)}_{\text{int}}.\label{SigPosy}
$
Hence, the inequality constraint~\eqref{INSR}, we obtain
\begin{align}
\frac{\psi_j\left( \Sigma^{(1)}_{j_\text{int}} + \Sigma_{j_{n_{\text{s}}}} + \Sigma_{j_{n_{\text{fc}}}}\right)}{\Sigma_{j_\text{des}}+\psi_j\Sigma^{(2)}_{j_\text{int}}}\leq 1,\ \forall j\in\mathcal{N}.\label{INSR1}
\end{align}
The left hand-side of the inequality constraint~\eqref{INSR1} is the division of posynomials, which can not be converted to a convex function. Problem~\eqref{A1} is a signomial program (SP)~\cite{Boyd2007}, which can be converted to a complementary geometric program (GP). This program allows upperbound constraint on the division of two posynomials. The denominator of~\eqref{INSR1} is approximated by a monomial function (known as condensation method~\cite{Chiang2005}) based on the following lower-bound
$
\sum_{k}c_k\mu_k\geq \prod_{k} \mu^{c_k}_k,
$
which states the relationship between arithmetic and geometric mean. This lower-bound on the denominator of~\eqref{INSR1} operates as an upper-bound on the whole expression. By defining $\hat{\mu}_k=c_k\mu_k$ we get
\begin{align}
\sum_{k}\hat{\mu}_k\geq \prod_{k} \left(\frac{\hat{\mu}_k}{c_k}\right)^{c_k}.\label{approx1}
\end{align}
Now, we utilize this inequality in the SINR constraints of the $j$th target in~\eqref{INSR1}. The denominator of~\eqref{INSR1} is rewritten as the summation of monomials by
\begin{align}
\Sigma_{j_D}=\Sigma_{j_\text{des}}+\psi_j\Sigma^{(2)}_{j_\text{int}}=
\sum_{k=1}^{K+K^{'}}\hat{\mu}_{jk},\label{SigmaD}
\end{align}
where $\hat{\mu}_{jk}$ are the individual monomials. Furthermore, $K^{'}$ is the number of monomials that the posynomial $\Sigma^{(2)}_{j_\text{int}}$ consists of, which can be quantified according to lemma 1. Then from~\eqref{approx1} and~\eqref{SigmaD},
\begin{align}
\Sigma_{j_D}=\sum_{k=1}^{K+K^{'}} \hat{\mu}_{jk}\geq \prod_{k=1}^{K+K^{'}} \left(\frac{\hat{\mu}_{jk}}{c_{jk}}\right)^{c_{jk}}=\tilde{\Sigma}_{j_D}(c_{jk}),\label{def2}
\end{align}
where $\tilde{\Sigma}_D$ is a function of $c_{jk}$, which need to be optimized to fulfill the inequality with equality. For that, $c_{jk}$ must be a function of $\alpha_k,\ \forall k$ as
$
c_{jk}=\frac{\hat{\mu}_{jk}}{\Sigma_{j_D}}\label{cjk}
$.
Due to the inter-dependency of the optimization parameters, we optimize $\alpha_k,\ \forall k$ and $c_{jk}$ successively in an iterative fashion. That means, $c_{jk},\ \forall j,k$ is optimized for the current iteration based on the solution of $\hat{\mu}_{jk}$ in the previous iteration. Notice that the lower-bound in~\eqref{def2} is the approximation of $\Sigma_D$ around any feasible $\alpha_k,\ \forall k$, though sub-optimal. Hence, by improving $\alpha_k\ \forall k$ and $p_j\ ,\forall j\in\mathcal{N}$ at each iteration, the approximation around it (defined by $c_{jk}$) is utilized for the next iteration. 
The convergence of the algorithm is numerically illustrated in section~\ref{NumRes}.

\section{Numerical Results}\label{NumRes}
In this section, we provide the simulation results for a two-target single-clutter environment, i.e., $N=2,N'=1$. The number of antennas at the transmitter is assumed to be 4, i.e., $M=2$ and $M^{'}=2$ (two antennas per dimension). The fusion center is equipped with 10 antennas, i.e., $R=10$. We assume 2 targets are at the azimuth and elevation angles
$
\boldsymbol{\theta}=[20\ 45],\ \boldsymbol{\phi}=[40\ 30]
$, respectively. \balance
Moreover, there exists a single clutter at the azimuth $\theta=70$ and elevation $\phi=85$.
Furthermore, the sensors maximum amplification factor is assumed to be equal to $2$, i.e., $\alpha_{\text{max}}=2$. The noise variance at the sensors and fusion center are assumed to be equal to 0.5, i.e., $\sigma^{2}_\text{fc}=\sigma^{2}_{n_k}=0.5,\ \forall k$. Considering MRC at the receiver, the sum-power minimization problem is solved iteratively under per-target SINR constraints. This problem is a signomial program, which is turned to a geometric program and solved iteratively until convergence. The convergence of the algorithm for joint transmit power and sensor amplification minimization problem is depicted in Fig.~\ref{Res:Converg1}, where we observe the fast convergence. Assuming maximum amplification at the sensors, the transmit power minimization problem is also a signomial problem, which is treated similarly. The minimum sum-power consumption for this case (maximum amplification) is compared to the case with optimal amplification in Fig.~\ref{Res:MRC}. With maximum amplification at the sensors, MRC is compared with ZF Fig.~\ref{Res:MRC}. In this figure, we observe that MRC is optimal when the SINR demands are sufficiently low. However, ZF outperforms MRC as the interference increases, hence, it is efficient to zero force the interference. Intuitively, by zero-forcing processing at the fusion center, interference-free signaling dimensions becomes less than the number of available dimensions $KR$. This is due to reserving $N+N'-1$ dimension for null steering. This leaves us with $KR-N-N'+1$ signaling dimensions. Therefore, comparing ZF and MRC we notice the trade-off between sacrificing some dimensions in expense of obtaining interference-free dimensions, and utilizing all dimensions. 
\bibliographystyle{IEEEtran}
\bibliography{reference}
\end{document}